\documentclass[12pt,seceqn]{elsart}

\usepackage{amsmath,amssymb,amsfonts}
\usepackage[mathcal]{euscript}
\usepackage{graphicx}
\usepackage{cite}

\newcommand{\comment}[1]{}

\newcommand{\ie}{\textit{i.e.}}

\newcommand{\mathnotation}[2]{\newcommand{#1}{\ensuremath{#2}}}

\renewcommand{\l}{\left}			
\renewcommand{\r}{\right}			
\mathnotation{\pd}{\partial}			
\mathnotation{\ee}{{\mathrm e}}			
\mathnotation{\ldef}{\mathrel{\raisebox{.069ex}{:}\!\!=}}
\mathnotation{\rdef}{\mathrel{=\!\!\raisebox{.069ex}{:}}}
\mathnotation{\levicivita}{\varepsilon}		

\mathnotation{\grad}{\nabla}			
\mathnotation{\curl}{\grad\times}		
\mathnotation{\lapl}{\nabla^2}			

\mathnotation{\third}{\textstyle\frac{1}{3}}

\mathnotation{\covd}{\grad}			
\mathnotation{\Dt}{\mathcal{D}}			
\mathnotation{\ptconnex}{\gamma}		
\mathnotation{\Connex}{\Gamma}			

\mathnotation{\dint}{\,{\mathrm{d}}}		
\mathnotation{\pheq}{&\phantom{=}}		
\renewcommand{\time}{t}				
\mathnotation{\lyapexp}{\lambda}		
\mathnotation{\lyapexpinf}{\lyapexp^\infty}	
\mathnotation{\x}{x}				
\mathnotation{\xv}{\x}				
\mathnotation{\velc}{v}				
\mathnotation{\vel}{\velc}			
\mathnotation{\flow}{\varphi}			
\mathnotation{\flowt}{\flow^\time}		
\mathnotation{\dotflowt}{\dot\flow^\time}
\mathnotation{\Manif}{{\mathcal{U}}}		
\mathnotation{\Tangent}{\mathrm{T}}		
\mathnotation{\Diff}{D}				
\mathnotation{\vvec}{X}				
\mathnotation{\wvec}{Y}				
\mathnotation{\G}{G}				
\mathnotation{\A}{A}				
\mathnotation{\M}{M}				
\mathnotation{\dM}{{\mathcal{M}}}		
\mathnotation{\Hesm}{K}				
\mathnotation{\sdim}{n}				
\mathnotation{\ns}{\sdim_{\mathrm{s}}}		
\mathnotation{\NI}{N_{\mathrm{I}}}		
\mathnotation{\NII}{N_{\mathrm{II}}}		

\mathnotation{\wdir}{w}				
\mathnotation{\wdirv}{\wdir}			

\mathnotation{\edir}{e}				
\mathnotation{\edirv}{\edir}			
\mathnotation{\ediru}{\hat{\edir}}		
\mathnotation{\ediruv}{\ediru}			
\mathnotation{\ediruinf}{\ediru^\infty}		
\mathnotation{\ediruvinf}{\ediruinf}		

\mathnotation{\lagrc}{a}			
\mathnotation{\lagrcv}{\lagrc}			
\mathnotation{\metrica}{g}			
\mathnotation{\metricx}{h}			
\mathnotation{\dsq}{k}				
\mathnotation{\detmetrica}{\det\metrica}	
\mathnotation{\nugr}{\Lambda}			
\mathnotation{\nudec}{\gamma}			
\newcommand{\norm}[1]{\l\|{#1}\r\|}		
\newcommand{\transp}[1]{{#1}^T}			

\mathnotation{\Q}{Q}				
\mathnotation{\V}{V}				
\mathnotation{\F}{F}				%
\mathnotation{\U}{U}
\mathnotation{\Vu}{\widehat\V}			

\mathnotation{\Vgrnu}{\Psi}			
\mathnotation{\VgrU}{\Phi}			
\mathnotation{\VgrV}{\Theta}			
\mathnotation{\X}{X}				
\mathnotation{\Rcurv}{R}			
\mathnotation{\Scurv}{S}			
\mathnotation{\Riccirotc}{\omega}		

\mathnotation{\Vgrnut}{\widetilde{\Vgrnu}}
\mathnotation{\VgrUt}{\widetilde{\VgrU}}
\mathnotation{\VgrVt}{\widetilde{\VgrV}}
\mathnotation{\Vgrnuinf}{\Vgrnu^\infty}
\mathnotation{\VgrVinf}{\VgrV^\infty}
\mathnotation{\homo}{{\mathrm{h}}}		
\mathnotation{\Vgrnudrive}{\Vgrnu^{\mathrm{drive}}}
\mathnotation{\VgrVdrive}{\VgrV^{\mathrm{drive}}}
\mathnotation{\Hesmt}{\widetilde{\Hesm}}
\mathnotation{\Vt}{\widetilde{\V}}
\mathnotation{\Vinf}{\V^\infty}

\mathnotation{\ppot}{\psi}			

\def\SVD{SVD}
\def\QR{$QR$}
\def\GK{Greene~\&~Kim}

\begin{document}

\begin{frontmatter}

\title{Differential Constraints in\\ Chaotic Flows on Curved Manifolds}
\author{Jean-Luc Thiffeault}

\address{Department of Applied Physics and Applied Mathematics\\
	Columbia University, New York, NY 10027, USA}

\ead{jeanluc@mailaps.org}

\begin{abstract}

The Lagrangian derivatives of finite-time Lyapunov exponents and the
corresponding characteristic directions are shown to satisfy time-asymptotic
differential constraints in chaotic flows.  The constraints are valid for any
metric tensor, and are realised with exponential accuracy in time.  Some of
these constraints were derived previously for chaotic systems on
low-dimensional Euclidean spaces, by requiring that the Riemann curvature
tensor vanish in Lagrangian coordinates.  The new derivation applies in any
number of dimensions, predicts the number of constraints for a given flow, and
provides a rigorous convergence rate of the constraints.

\end{abstract}

\end{frontmatter}

\section{Introduction}
\label{sec:intro}

A deep understanding of chaotic systems originates from the study of
time-asymptotic dynamics~\cite{Eckmann1985}.  The Lyapunov exponents (or
characteristic exponents) are the most well-known instance; they describe the
average rate of exponential separation of neighbouring trajectories, in the
infinite-time limit~\cite{Oseledec1968,Shimada1979,Benettin1978}.  The
Lyapunov exponents are often used as a criterion for chaos: on a bounded
domain, at least one positive exponent is required.  They have proven quite
rich, for example in characterising the fractal dimension of attractors in
dissipative systems~\cite{Farmer1983,Sommerer1993}.

The convergence of the Lyapunov exponents is extremely slow, typically
logarithmic in time.  This is due to the fact that they are inherently
nonlocal and require the fiducial trajectory (the trajectory along which the
exponents are evaluated) to explore the attractor (or invariant region, for
nonattracting systems) very thoroughly, as the greatest stretching often comes
from rare excursions near hyperbolic fixed points.  It is thus debatable
whether the exponents are relevant to the finite-time dynamics of a single
trajectory, or to an ensemble of tightly bunched trajectories, since a
laboratory or numerical experiment will usually have ended long before the
exponents have converged.

Hence the shift of emphasis to the finite-time Lyapunov exponents (FTLEs).
These are the same as the Lyapunov exponents (\ie, the infinite-time Lyapunov
exponents) but averaged over a finite time, as the name implies.  On short to
moderate time scales, the FTLEs do not show convergence at all (that is, the
average rate of separation is not exponential), but they are still connected
to a rapid, quasi-exponential average separation of neighbouring trajectories.

The rapid separation of trajectories leads to a quasi-exponential convergence
of the characteristic directions, the directions in the space of initial
conditions (Lagrangian coordinates, which label fluid elements) exhibiting a
given characteristic rate of separation.  Thus, even though the FTLEs have not
usually converged, on moderate time scales the characteristic directions
converge very rapidly~\cite{Goldhirsch1987}.  This is because the
characteristic directions are local quantities, and they will typically
converge after the fiducial trajectory has explored a portion of their
neighbourhood.

In this paper we show that there are constraints on the derivatives of FTLEs
and of characteristic directions.  The derivatives are taken with respect to
the initial starting coordinate of the fiducial trajectory---the Lagrangian
co\-or\-di\-nate---and so are called Lagrangian derivatives.  The differential
constraints are asymptotic, in the sense that they are satisfied with
quasi-exponential accuracy in time.

For two- and three-dimensional dynamical systems, these constraints were first
derived by Tang and Boozer~\cite{Tang1996} and Thiffeault and
Boozer~\cite{Thiffeault2001}.  The method used there was to transform the
Euclidean metric defined on the space in which the motion takes place
(Eulerian coordinates) to the space of initial conditions (Lagrangian
coordinates, or fluid labels).  The metric in Lagrangian coordinates exhibits
quasi-exponential blowup, but its associated Riemann curvature tensor must
vanish identically, since the underlying space is flat.  Thus, by balancing
the various terms in the Riemann curvature tensor, it was shown that the
curvature can only vanish identically if the FTLEs and characteristic
directions obey asymptotic differential constraints.  These constraints were
also confirmed numerically for discrete maps~\cite{Tang1996,Tang1999b} and for
flows~\cite{Thiffeault2001,Thiffeault2001b}.

Here we propose a different approach, with an emphasis on flows.  By
performing a singular value decomposition (\SVD) of the tangent map of the
flow (the Jacobian of the transformation from Eulerian to Lagrangian
coordinates), the quasi-exponentially growing terms can be evolved
independently from each other, so that their different growth rates do not
pose a problem numerically.  This amounts to a reorthonormalisation scheme,
where the dominant eigenvectors are subtracted from the subdominant
ones~\cite{Wolf1985,Goldhirsch1987,Geist1990}.  Greene and
Kim~\cite{Greene1987,Greene1989} devised a numerical scheme to evolve the
components of the \SVD\ continuously.  Thiffeault~\cite{Thiffeault2001b}
extended their method to compute the Lagrangian derivatives of FTLEs and
characteristic eigenvectors.  This amounts to a numerical scheme for
calculating the Hessian (the tensor of second partial derivatives) of the
transformation from Eulerian (fixed or laboratory coordinates) to Lagrangian
(fluid element labels).  Related work on maps was done by Dressler and
Framer~\cite{Dressler1992} and Taylor~\cite{Taylor1993}.  Whereas the Jacobian
contains information on the deformation of an infinitesimal ball of initial
conditions into an ellipsoid, the Hessian describes higher-order (nonlinear)
deformations.

The constraints are obtained in two steps.  First, the time-asymptotic
behaviour of Lagrangian derivatives is derived by analysing the equations for
the components of the \SVD, in a manner analogous to
Ref.~\cite{Thiffeault2001b} but allowing for a nontrivial (\ie, non-Euclidean)
metric.  Second, the symmetry of the Hessian is used, and Eulerian quantities
are replaced by their asymptotic form.  What is left are expressions
generalising the constraints of Ref.~\cite{Thiffeault2001} to an arbitrary
number of dimensions and choice of metric.  The new method of derivation also
predicts the convergence rate of the constraints, something that was missing
from earlier analyses.

Note that even though it is well-known that the Lyapunov exponents are
independent of the particular positive-definite metric chosen to measure
them~\cite{Oseledec1968,Greene1989}, this is not true of the finite-time
exponents and the characteristic directions.  The details of the metric may
affect the short- to moderate-time behaviour of the FTLEs greatly, and the
characteristic directions are different even in the infinite-time limit.
Hence, the derivation of the present results for the case of an arbitrary
metric is relevant from a mathematical standpoint.

From a physical perspective, metrics that differ from the ordinary Euclidean
case often arise.  Examples are the advection--diffusion equation with an
an\-isotropic, inhomogeneous diffusion tensor, and two-dimensional flow on a
deformable membrane~\cite{Aris,Thiffeault2001f}.  Even when the underlying
space is Euclidean, there are many cases where it is advantageous to use
coordinates better suited to the problem at hand, which lead to a nontrivial
metric (spherical, toroidal, rotating, etc.).  Hence, there is also physical
motivation in keeping the derivation of the constraints as general as possible
by considering an arbitrary, time- and space-dependent metric.

The outline of this paper is as follows.  In Section~\ref{sec:curvedsp} we
review the mathematical description of a continuous dynamical system (flow) on
a manifold with curvature.  We define the Jacobian and the Hessian of the
flow, and describe the metric on the Eulerian and Lagrangian tangent spaces of
the manifold.  We also choose a convenient covariant time derivative.  In
Section~\ref{sec:orthonormal-frames} we define the orthonormal frames that
will allow the separation of the different scales of the system along
characteristic directions.  This leads to the singular value decomposition and
the time evolution equations for the orthonormal frames.  From these we can
define the finite-time Lyapunov exponents and characteristic directions.  We
devote Section~\ref{sec:LagrDerivs} to showing that the nontrivial metric does
not significantly affect the asymptotic forms of the Lagrangian derivatives
derived in Ref.~\cite{Thiffeault2001b}, allowing us to make use of the results
therein.  Section~\ref{sec:constraints} contains the main results of the
paper: we derive two types of constraints, and give the number of constraints
of each type as a function of dimension and number of contracting directions
(negative Lyapunov exponents).  In Section~\ref{sec:curvature}, we make a
connexion to the previous method (the curvature method of
Refs.~\cite{Tang1996,Thiffeault2001}) of finding constraints.  Finally, we
offer a summary and discussion of the implications of the results of the paper
in Section~\ref{sec:discussion}.

\section{Chaotic Dynamics in a Curved Space}
\label{sec:curvedsp}

In this section we introduce notation and review a number of concepts
necessary for the remainder of the paper.  In
Section~\ref{sec:dynam-syst-manif} we discuss a general dynamical system
defined on a manifold.  In Section~\ref{sec:metric} we equip that manifold
with a metric tensor.  To assist in writing equations in manifestly covariant
form, in Section~\ref{sec:covar-time-deriv} we introduce a covariant form of
the time derivative.

\subsection{Dynamical Systems on a Manifold}
\label{sec:dynam-syst-manif}

We consider the dynamical system on a smooth compact manifold~$\Manif$
\begin{equation}
	\dot\xv = \vel(\time,\xv),
	\label{eq:dynsys}
\end{equation}
where the overdot indicates a derivative with respect to time.
Here~$\xv=\xv(\time)\in\Manif$ is a smooth function of~$\time\in\Rset$,
and~$\vel:\Manif\rightarrow\Tangent\,{\Manif}$ is a smooth vector field,
with~$\Tangent\,{\Manif}$ the tangent bundle of~$\Manif$.  We say that~$\vel$
generates a flow~$\flowt: \Manif \rightarrow \Manif$, where~$\flowt(\lagrcv)$
is a diffeomorphism defined for all~$\lagrcv$ in~$\Manif$ and~$\time\in\Rset$.
The flow~$\flowt$ exists, and it satisfies Eq.~\eqref{eq:dynsys} if we
set~\hbox{$\xv = \flowt(\lagrcv)$}~\cite[p.~304]{ArnoldODE}.  We
choose~$\flowt$ such that~\hbox{$\flow^0(\lagrcv)=\lagrcv$}, defining~$\lagrcv$
as the initial condition of a trajectory.  By analogy with fluids, the~$\xv$
are called Eulerian coordinates and the~$\lagrcv$ Lagrangian coordinates.

The derivative (also known as the tangent map, Jacobian, or push-forward)
at~$\lagrcv$ of~$\flowt$ is denoted~\hbox{$\flowt_{*\lagrcv}:
\Tangent_{\lagrcv}\Manif \rightarrow \Tangent_{\flowt(\lagrcv)}\Manif$}, and
satisfies the equation
\begin{equation}
	\dotflowt_{*\lagrcv} = \Diff\vel(\time,\flowt(\lagrcv))
	\cdot\flowt_{*\lagrcv}\,,
	\qquad \flow^0_{*\lagrcv} = \mathrm{Id},
	\label{eq:TangentODE}
\end{equation}
obtained by differentiating Eq.~\eqref{eq:dynsys}.  We use a dot, as on the
right-hand side of Eq.~\eqref{eq:TangentODE}, to indicate the action of a
linear map.  In a particular chart, Eq.~\eqref{eq:TangentODE} can be written
in coordinate form as
\begin{equation}
	{(\dotflowt_{*\lagrcv})^i}_q = \frac{\pd\vel^i}{\pd\xv^k}\,
		{(\flowt_{*\lagrcv})^k}_q\,,
	\qquad {(\flow^0_{*\lagrcv})^i}_q = {\delta^i}_q,
	\label{eq:TangentODEcoord}
\end{equation}
where~${\pd\vel^i}/{\pd\xv^k}$ is evaluated at~$\xv = \flowt(\lagrcv)$,
and we are using the Einstein convention of summing over repeated indices
from~$1$ to~$\sdim$, where~$\sdim$ is the dimension of~$\Manif$.

Equation~\eqref{eq:TangentODEcoord} introduces a convention we shall use for
the remainder of the paper: the indices $i,j,k,\ell$ denote vectors in
Eulerian space ($\Tangent_{\xv}\Manif$), while the indices $p,q,r$ denote
vectors in Lagrangian space ($\Tangent_{\lagrcv}\Manif$).
Since~$\flowt_{*\lagrcv}$ is a map from~$\Tangent_{\lagrcv}\Manif$
to~$\Tangent_{\xv}\Manif$, we write its components
as~${(\flowt_{*\lagrcv})^k}_q$.

If we imagine an infinitesimally small ``ball'' of points distributed around
the initial condition~$\lagrcv$ at~$\time=0$, then the tangent
map~$\flowt_{*\lagrcv}$ captures the deformation of that ball into an
ellipsoid under the action of the flow~$\flowt$ (see Ref.~\cite{Geist1990} for
a more detailed description).  For a chaotic flow, at least one axis of this
ellipsoid grows exponentially in time, the growth rate being characterised by
the largest Lyapunov exponent~\cite{Oseledec1968,Eckmann1985} (for finite
times the growth rate is only roughly exponential, or quasi-exponential, but
from now on we drop the quasi- prefix).  This renders the direct numerical
evaluation of Eq.~\eqref{eq:TangentODEcoord} for moderately large times all
but impossible due to roundoff error: all the columns
of~${(\flowt_{*\lagrcv})^i}_q$ become aligned with the eigendirection of
fastest stretching.  A suitable matrix decomposition is needed to separate the
growth of the different axes of the ellipsoid such that the fastest expanding
direction does not overwhelm the others, that is, a reorthonormalisation
scheme~\cite{Wolf1985,Goldhirsch1987,Geist1990}.  In Section~\ref{sec:decomp}
we achieve this using the singular value decomposition (\SVD), in a manner
analogous to \GK~\cite{Greene1987,Greene1989}.


In this paper, as in earlier work~\cite{Thiffeault2001b}, we are interested in
deformations of a ball of initial conditions beyond ellipsoidal, so we want to
obtain higher derivatives of~$\flowt$.  The second derivative of~$\flowt$ may
be regarded as a symmetric bilinear form \hbox{$\flowt_{**\lagrcv}:
\Tangent_{\lagrcv}\Manif\times\Tangent_{\lagrcv}\Manif \rightarrow
\Tangent_{\flowt(\lagrcv)}\Manif$}, satisfying
\begin{equation}
	\dotflowt_{**\lagrcv} = \Diff\vel(\time,\flowt(\lagrcv))
		\cdot\flowt_{**\lagrcv}
	+ \Diff^2\vel(\time,\flowt(\lagrcv))[
		\flowt_{*\lagrcv}\otimes\flowt_{*\lagrcv}]\,,
	\qquad \flow^0_{**\lagrcv} = 0.
	\label{eq:HessianODE}
\end{equation}
(See Dressler and Farmer~\cite{Dressler1992} for a similar derivation for
maps.)  Writing out the components on a particular chart explicitly,
Eq.~\eqref{eq:HessianODE} has the form
\begin{equation}
	(\dotflowt_{**\lagrcv})^i_{pq}
	= \frac{\pd\vel^i}{\pd\xv^k}\,
	(\flowt_{**\lagrcv})^k_{pq}
	+ \frac{\pd^2\vel^i}{\pd\xv^k\pd\xv^\ell}\,
	{(\flowt_{*\lagrcv})^k}_p\,{(\flowt_{*\lagrcv})^\ell}_q\,,
	\qquad (\flow^0_{**\lagrcv})^i_{pq} = 0.
	\label{eq:HessianODEcoord}
\end{equation}
We call the quadratic form~$\flowt_{**\lagrcv}$ the Hessian of~$\flowt$.  The
growth rate of the Hessian is characterised by \emph{generalised} or
\emph{higher-order} Lyapunov
exponents~\cite{Dressler1992,Taylor1993,Thiffeault2001b}.  These are typically
larger in magnitude than the Lyapunov exponents associated with the growth of
the tangent map~$\flowt_{*\lagrcv}$ (the Jacobian).  This means that
numerically integrating Eq.~\eqref{eq:HessianODEcoord} directly is even more
problematic than for the tangent map.  In Ref.~\cite{Thiffeault2001b}
extensions of the \SVD\ and \QR\ (orthogonal-triangular) decomposition methods
were introduced to resolve this numerical difficulty.  This allowed the
precise verification of the differential constraints predicted in earlier
work~\cite{Tang1996,Thiffeault2001}.

\subsection{The Metric}
\label{sec:metric}

To measure the norm of vectors, we now introduce a Riemannian
metric~$(\cdot,\cdot)_\xv: \Tangent_{\xv}\Manif \times \Tangent_{\xv}\Manif
\rightarrow \Rset$.  This induces a corresponding
metric~$(\cdot,\cdot)^\time_\lagrcv$ on~$\Tangent_{\lagrcv}\Manif$ defined by
\begin{equation}
	(\vvec\,,\wvec)^\time_\lagrcv \ldef
	(\flowt_{*\lagrcv}\cdot\vvec\,,
	\flowt_{*\lagrcv}\cdot\wvec)_\xv\,,
	\label{eq:metriclagrc}
\end{equation}
for all~$\vvec,\wvec\in\Tangent_{\lagrcv}\Manif$.  At~$\time=0$ the two
metrics coincide.  The reason why we introduce a metric
on~$\Tangent_{\xv}\Manif$ and from it obtain the metric
on~$\Tangent_{\lagrcv}\Manif$, and not the other way around, is best
understood in terms of the fluid analogy.  We wish to measure the growth of
the vector~$\vvec$ as it is ``dragged'' by the flow.  Initially, the norm
of~$\vvec$ is~$(\vvec,\vvec)_\xv^{1/2}$, and at a later time~$\time$ its norm
is~$(\flowt_{*\lagrcv}\cdot\vvec\,, \flowt_{*\lagrcv}\cdot\vvec)_\xv^{1/2}$.
The latter is then used in Eq.~\eqref{eq:metriclagrc} to define a
time-dependent metric directly on~$\Tangent_{\lagrcv}\Manif$, with no explicit
reference to~$\flowt$.  Thus, it is the metric on~$\Tangent_{\xv}\Manif$ that
is given.

In component form, we have~\hbox{$(\vvec\,,\wvec)^\time_\lagrcv =
\metrica_{pq}(\time,\lagrcv)\,\vvec^p\,\wvec^q$}, with
\begin{equation}
	\metrica_{pq}(\time,\lagrcv)
	= {(\flowt_{*\lagrcv})^i}_p\,\metricx_{ij}
		(\time,\flowt(\lagrcv))\,
		{(\flowt_{*\lagrcv})^j}_q\,,
	\label{eq:metricacomp}
\end{equation}
where~$\metricx_{ij}(\time,\flowt(\lagrcv))$ are the components of the metric
on~$\Tangent_{\xv}\Manif$.  In Euclidean space, we
have~$\metricx_{ij}(\time,\flowt(\lagrcv)) = \delta_{ij}$, but we shall retain
the more general form.  Both~$\metrica_{pq}$ and~$\metricx_{ij}$ are
positive-definite symmetric matrices, since we chose a proper
(positive-definite) Riemannian metric.

\subsection{A Covariant Time Derivative}
\label{sec:covar-time-deriv}

To ensure that we are always dealing with tensors---and thus guarantee the
covariance of all expressions---it is convenient to introduce a covariant
version of the time-derivative operator~\cite{Thiffeault2001e}.  In the
Eulerian basis, the components of this operator~$\Dt$ acting on a
vector~$\vvec\in\Tangent_{\xv}\Manif$ are
\begin{equation}
	\Dt{\vvec^i}
	\ldef \dot\vvec^i
		+ \Connex^i_{k\ell}\,\vvec^k\vel^\ell
		+ {\ptconnex^i}_k\,\vvec^k\,,
	\label{eq:Dtdef}
\end{equation}
where the Riemann--Christoffel connexions are defined as
\begin{equation}
	\Connex^k_{ij} \ldef \half\metricx^{k\ell}\l(
	\metricx_{i\ell,j} + \metricx_{j\ell,i}
	- \metricx_{ij,\ell}\r),
\end{equation}
and
\begin{equation}
	{\ptconnex^i}_j \ldef
	\half\metricx^{i\ell}\,\frac{\pd\metricx_{\ell j}}{\pd\time}\,.
\end{equation}
We use the symbol~$\covd_j$ to denote a covariant derivative with respect
to~$\xv^j$,
\begin{equation}
	\covd_j\vvec^i \ldef \frac{\pd\vvec^i}{\pd\x^j}
		+ \Connex^i_{jk}\,\vvec^k.
\end{equation}

Without the~$\ptconnex$ term, Eq.~\eqref{eq:Dtdef} is the usual definition of
covariant differentiation along a curve~\cite{Wald}, in our case the curve
being the trajectory of~$\xv$ with tangent vector~$\vel$.  The extra term
allows the metric to depend explicitly on time whilst preserving the
compatibility property~\hbox{$\Dt{\metricx_{ij}}=0$}.  Note that~$\Connex$,
$\ptconnex$, and $\vel$ do not transform as tensors, but their combination
defined by Eq.~\eqref{eq:Dtdef} does~\cite{Thiffeault2001e}.

\section{Two Orthonormal Frames}
\label{sec:orthonormal-frames}

In Section~\ref{sec:dynam-syst-manif}, we mentioned that direct numerical
integration of the evolution equation~\eqref{eq:TangentODEcoord} for the
tangent map~$\flowt_{*\lagrcv}$ is impractical due to the dominance of the
eigenvector associated with the direction of fastest stretching.  In the
present section we use an appropriate matrix decomposition to separates the
vastly different scales contained in~$\flowt_{*\lagrcv}$
(Section~\ref{sec:decomp}), as done by \GK~\cite{Greene1987,Greene1989}, but
our treatment includes a metric with explicit time-dependence.  We aim to
write the equations of motion for the matrices of the decomposition in
manifestly covariant form (Section~\ref{sec:time-evolution}), using the
covariant time derivative of Section~\ref{sec:covar-time-deriv}.  In addition
to clarifying the r\^{o}le of a nontrivial metric in chaotic dynamics, the
covariant form allows a more straightforward derivation of the equations of
motion for the Lagrangian derivatives, as we shall see in
Section~\ref{sec:LagrDerivs}.  Finally, in Section~\ref{sec:lyapunov-exp} the
finite-time Lyapunov exponents and characteristic directions are defined.

\subsection{Decomposition of $\flowt_{*\lagrcv}$ into Orthonormal Bases}
\label{sec:decomp}

The tangent map~$\flowt_{*\lagrcv}$ is the Jacobian~$\pd\xv/\pd\lagrcv$ of the
transformation from Lagrangian coordinates ($\lagrcv$) to Eulerian coordinates
($\xv$).  Hence, it tells us how an set of orthonormal vectors (with respect
to the metric~$\metricx$) at~$\lagrcv$ is transformed to a set of vectors
at~$\xv$.  This last set is not orthonormal in general.

We can however decompose~$\flowt_{*\lagrc}$ into an outer product of
orthonormal bases,
\begin{equation}
	{(\flowt_{*\lagrc})^i}_q =
	\sum_{\sigma,\tau}
		(\wdirv_\sigma)^i\,(\edirv_\tau)_q\,\eta^{\sigma\tau}.
	\label{eq:flowtdecomp}
\end{equation}
where~$\eta^{\sigma\tau} = \delta^{\sigma\tau}$, since we have a
positive-definite metric, but we write it as~$\eta$ following standard
notation~\cite{Wald}.  Note that for Greek indices we will \emph{not} use the
Einstein sum convention; all sums will be written out explicitly.  By
``orthonormal,'' we mean that the bases~$\wdir_\sigma$ and~$\edir_\sigma$
satisfy
\begin{equation}
	\metricx_{ij}(\time,\xv) = \sum_{\sigma,\tau}
		(\wdirv_\sigma)^i\,(\wdirv_\tau)^j\,\eta^{\sigma\tau},
	\qquad
	\metrica_{pq}(\time,\lagrcv) = \sum_{\sigma,\tau}
		(\edirv_\sigma)_p\,(\edirv_\tau)_q\,\eta^{\sigma\tau}.
	\label{eq:metricdiag}
\end{equation}
The orthonormal basis vectors~$\edirv_\sigma$
depend on~$\lagrcv$ and~$\time$, and the basis vectors~$\wdirv_\tau$ depend
on~$\xv$ and~$\time$; they diagonalise~$\metrica$ and~$\metricx$,
respectively.  The basis~$\{\edirv_\sigma\}$ is a noncoordinate basis, \ie, it
does not correspond to the natural tangent vectors~$\{\pd/\pd\lagrcv^q\}$ of
the coordinate system~$\lagrcv$, or of any other coordinate system.  The same
is true of the basis~$\{\wdirv_\sigma\}$.
It follows from Eq.~\eqref{eq:metricdiag} that
\begin{equation}
	(\wdirv_\tau)^\ell\,(\wdirv_\sigma)_\ell = \eta_{\tau\sigma}\,,
	\qquad
	(\edirv_\tau)^q\,(\edirv_\sigma)_q = \eta_{\tau\sigma}\,.
	\label{eq:wedirortho}
\end{equation}

In the appendix we show that, using the standard singular value decomposition
(\SVD), we can always write
\begin{equation}
	{\M^i}_q = \sum_{\sigma,\tau}{\U^i}_\sigma\,
		\F^{\sigma\tau}\,\V_{q\tau}\,,
	\label{eq:decomp}
\end{equation}
where the matrix~$\M$ denotes the components of~$\flowt_{*\lagrcv}$ in a
particular chart.  The notation for the components of the decomposition is
that of Refs.~\cite{Greene1987,Greene1989,Thiffeault2001b}.  The matrix~$\F$
is diagonal, and the matrices~$\U$ and $\V$ satisfy the orthogonality
relations,
\begin{equation}
	\metricx_{ij}\,{\U^i}_\sigma\,{\U^j}_\tau
		= \eta_{\sigma\tau}, \qquad
	\eta^{pq}\,\V_{p\sigma}\,\V_{q\tau} = \eta_{\sigma\tau}.
\end{equation}
Note that~$\U$ is orthogonal with respect to the metric
on~$\Tangent_{\xv}\Manif$, but~$\V$ is orthogonal with respect to~$\eta^{pq}$,
which is \emph{not} the metric on~$\Tangent_{\lagrcv}\Manif$.  The reason for
this is that we want to avoid including in~$\V$ the exponentially growing
terms contained in~$\metrica$.

The decomposition~\eqref{eq:decomp} is unique up to permutations of rows and
columns.  The diagonal elements~$\F^{\sigma\sigma}\rdef\nugr^{\sigma}$ are
called the singular values of~$\M$.  Requiring that the singular values be
ordered decreasing in size makes the decomposition unique (for nondegenerate
eigenvalues).  We refer the reader to Refs.~\cite{Geist1990,Thiffeault2001b}
for a discussion of the geometrical significance of the \SVD\ in dynamical
systems.

From Eqs.~\eqref{eq:flowtdecomp} and~\eqref{eq:decomp}, we can make the
identification
\begin{equation}
	(\edirv_\sigma)_p = \V_{p\rho}\,{\F^\rho}_\sigma\,,
	\qquad
	(\edirv_\sigma)^p = {\Vu^p}{}_\rho\,
		{({\F^{-1}})^\rho}_\sigma\,,
	\label{eq:edirdef}
\end{equation}
where we used~$\eta_{\mu\nu}$ as a ``metric'' to raise and lower Greek
indices, and we defined~\hbox{${\Vu^p}{}_\rho \ldef \eta^{pq}\,\V_{q\rho}$} to
abridge the notation.  Note that~${\Vu^p}{}_\rho$ is a tensor, but is not
equal to~${\V^p}_\rho = \metrica^{pq}\,\V_{q\rho}$ because~$\eta^{pq}$ is not
the metric on~$\Tangent_{\lagrcv}\Manif$; hence the need for a different
symbol.

Analogously to~$\{\edirv_\sigma\}$, we can take the basis~$\{\wdir_\sigma\}$
to be
\begin{equation}
	(\wdirv_\sigma)^i = {\U^i}_\tau\,,
	\qquad
	(\wdirv_\sigma)_i = \U_{i\tau}\,.
	\label{eq:wdirdef}
\end{equation}
With the definitions~\eqref{eq:edirdef} and~~\eqref{eq:wdirdef} it is easy to
check that~$\{\edirv_\sigma\}$ and~$\{\wdirv_\sigma\}$ satisfy the
orthonormality conditions~\eqref{eq:metricdiag} and~\eqref{eq:wedirortho}.

\subsection{Time Evolution}
\label{sec:time-evolution}

\GK~\cite{Greene1987} derived equations of motion for~$\U$, $\F$, and~$\V$ for
a metric without explicit time-dependence.  Our derivation parallels theirs,
except for the use of the covariant time derivative of
Section~\ref{sec:covar-time-deriv} to allow for time-dependence of the
metric.

We differentiate the decomposition~\eqref{eq:decomp} using the
covariant derivative~\eqref{eq:Dtdef},
\begin{equation}
	\Dt{\M^i}_q = 0 = \sum_{\sigma,\tau}\l[
		\Dt{\U^i}_{\sigma}\,\F^{\sigma\tau}\,\V_{q\tau}
		+ {\U^i}_{\sigma}\,\Dt\F^{\sigma\tau}\,\V_{q\tau}
		+ {\U^i}_{\sigma}\,\F^{\sigma\tau}\,\Dt\V_{q\tau}\r],
\end{equation}
and then contract with~$\U_{i\mu}\,{\Vu^q}{}_\nu$,
\begin{equation}
	\sum_{\sigma,\tau}\l[
	\U_{i\mu}\,\Dt{\U^i}_{\sigma}\,\F^{\sigma\tau}\,\eta_{\tau\nu}
		+ \eta_{\mu\sigma}\,\dot\F^{\sigma\tau}\,\eta_{\tau\nu}
		+ \eta_{\mu\sigma}\,\F^{\sigma\tau}\,
			{\Vu^q}{}_\nu\,\Dt\V_{q\tau}
	\r] = 0,
\end{equation}
where because~$\F^{\sigma\tau}$ is a scalar we replaced the covariant time
derivative by an ordinary one.  To simplify the notation, we use the
diagonality of~$\F$, and raise and lower Greek indices with~$\eta$, yielding
\begin{equation}
	\U_{i\mu}\,\Dt{\U^i}_{\nu}\,{\F^{\nu}}_{\nu}
		+ \dot\F_{\mu\nu}
		+ {\F^{\mu}}_{\mu}\,{\Vu^q}{}_\nu\,\Dt\V_{q\mu} = 0.
\end{equation}
The scalar~$\U_{i\mu}\,\Dt{\U^i}_{\nu}$ is antisymmetric in~$\mu$ and~$\nu$,
but not~${\Vu^q}{}_\nu\,\Dt\V_{q\mu}$, because~$\Dt\eta^{pq}\ne0$
since~$\eta^{pq}$ is not the metric.  However, the
quantity~{${\Vu^q}{}_\nu\,\dot\V_{q\mu} =
\eta^{pq}\,\V_{p\nu}\,\dot\V_{q\mu}$} is antisymmetric in~$\mu$ and~$\nu$, so
we expand the covariant derivative of~$\V$, \comment{The form of the covariant
derivative that we wrote down in Section \ref{sec:covar-time-deriv} is not
valid in Lagrangian coordinates.  See \texttt{covder.tex v1.28}, Eq.~(24).
First write down the derivative in terms of an arbitrary
connexion~$\alpha_{pq}$, then use the transformation law to arbitrary
coordinates.  Specialise to Eulerian to obtain the result below.}
\begin{equation}
	\U_{i\mu}\,\Dt{\U^i}_{\nu}\,{\F^{\nu}}_{\nu}
		+ \dot\F_{\mu\nu}
		+ {\F^{\mu}}_{\mu}\,{\Vu^q}{}_\nu
	\dot\V_{q\mu} - {\F^\nu}_\nu\,\G_{\mu\nu} = 0,
	\label{eq:solveforSVDdt}
\end{equation}
where
\begin{equation}
	\G_{\mu\nu} \ldef {\U^i}_\mu\,{\U^j}_\nu
		\l[\covd_j\vel_i + \ptconnex_{ij}\r],
	\label{eq:Gdef}
\end{equation}
and we used the transformation law from Lagrangian to Eulerian coordinates
imposed by the covariance of~$\Dt$~\cite{Thiffeault2001e}.  The matrix~$\G$ is
a covariant generalisation of the velocity gradient tensor, expressed in
the~$\U$ basis.  The~$\ptconnex$ term denotes straining due to the
time-dependence of the metric.

\comment{
From \texttt{covder.tex v1.28}, Eq. (25):
\begin{equation}
g_{ac}\,\grad_b V^c + \kappa_{ab} = \gamma^{\mathrm{ros}}_{ab} + \omega_{ab}
\end{equation}
where $\gamma^{\mathrm{ros}}$ is the rate-of-strain tensor, $\omega$ is the
vorticity tensor, and $\kappa$ is the coordinate rate-of-strain tensor.  $V$
is the velocity tensor and is defined as $\vel - \pd z/\pd\time$.  In the
Eulerian basis, we have $\pd\xv/\pd\time=0$, so
\begin{equation}
g_{i\ell}\,\grad_j \vel^\ell + \ptconnex_{ij} = \gamma^{\mathrm{ros}}_{ij} +
\omega_{ij}\,,
\end{equation}
which reduces to the above.  Hence, expect that the general form of~$\G$ is
\begin{equation}
	\G_{\mu\nu} = {\U^a}_\mu\,{\U^b}_\nu
		\l[\covd_b V_a + \kappa_{ab}\r]
	= {\U^a}_\mu\,{\U^b}_\nu
		\l[\gamma^{\mathrm{ros}}_{ab} + \omega_{ab}\r].
\end{equation}
This is very nice.  To prove that this is the general form, need to transform
from Lagrangian to a general coordinate system (did this on paper).  }

We can now solve Eq.~\eqref{eq:solveforSVDdt} for the various time derivatives
by using the antisymmetry of~$\U_{\ell\mu}\,\Dt{\U^\ell}_\nu$
and~${\Vu^q}{}_\mu\,{\dot\V}_{q\nu}$.  We obtain finally
\begin{align}
	{\dot\F^\mu}{}_\mu &=
		{\G^\mu}_\mu\,{\F^\mu}_\mu,\label{eq:SVDnugrODE}\\[5pt]
	\U_{\ell\mu}\,\Dt{\U^\ell}_\nu &= -\frac{\G_{\mu\nu}({\F^\nu}_\nu)^2
		+ \G_{\nu\mu}({\F^\mu}_\mu)^2}
		{({\F^\mu}_\mu)^2 - ({\F^\nu}_\nu)^2},
		&&\text{for $\mu \ne \nu$;}
	\label{eq:SVDUODE}\\[5pt]
	{\Vu^q}{}_\mu\,{\dot\V}_{q\nu}
		&= -\frac{{\F^\mu}_\mu\,{\F^\nu}_\nu}
		{({\F^\mu}_\mu)^2 - ({\F^\nu}_\nu)^2}\,\,\A_{\mu\nu},
		&& \text{for $\mu \ne \nu$;}
	\label{eq:SVDVODE}
\end{align}
where~$\A_{\mu\nu} \ldef \G_{\mu\nu} + \G_{\nu\mu}$.  These equations are
identical in form to those of \GK~\cite{Greene1987,Greene1989}, except for the
definition of~$\G$ and the use of the operator~$\Dt$ instead of the time
derivative in the~$\U$ equation.  Note that the matrix~$\A$ is the
rate-of-strain tensor (up to a possible factor of two, depending on the
convention used), expressed in the basis~$\U$.

The ordinary time derivative is used for~$\F$ because it is a scalar.
For~$\U$, we need to use the modified version~$\Dt$ of the time derivative to
take into account the nontrivial, possibly time-dependent metric.  The
ordinary derivative is used for the~$\V$ equation, even though it is a vector,
because we have effectively rescaled the metric~$\metrica_{pq}$ to
give~$\eta_{pq}$.  The reason for doing so is that for a chaotic flow the
elements of~$\metrica_{pq}$ grow exponentially, rendering the metric difficult
to use directly.  Instead, we have absorbed the exponential growth into
the~${\F^\mu}_\mu$, which are the coefficients of expansion of the
flow~\cite{Oseledec1968,Eckmann1985}.

The techniques used in this section also apply to the \QR\ method for
obtaining the Lyapunov exponents and characteristic
directions~\cite{Geist1990,Goldhirsch1987}.  One simply replaces the ordinary
time derivative of the~$Q$ equation by a covariant derivative (analogous to
the~$\U$), and the time derivatives of the elements of~$R$ are unchanged
(analogous to~$\F$ and~$\V$).  The~$\pd\vel/\pd\x$ term must be modified to
use the covariant spatial derivative, and a time derivative of the metric must
be included, exactly as in Eq.~\eqref{eq:Gdef}.  The \QR\ method has the
advantage of having no singularity in its equations of motion, as opposed to
Eqs.~\eqref{eq:SVDUODE} and~\eqref{eq:SVDVODE} which are singular
for~$\nugr_\mu=\nugr_\nu$.  However, the type of asymptotic analysis we do
here is better done with the \SVD\ method, as it was in
Ref.~\cite{Thiffeault2001b}.

\subsection{Lyapunov Exponents and Characteristic Directions}
\label{sec:lyapunov-exp}

The relation~\eqref{eq:edirdef} between the orthonormal
frame~$\{\edirv_\sigma\}$ and the \SVD\ matrices allows the definition of the
finite-time Lyapunov exponents
\begin{equation}
	\lyapexp_\sigma(\time,\lagrcv)
		\ldef \frac{1}{\time}\,\log\nugr_\sigma(\time,\lagrcv),
	\label{eq:FTLEdef}
\end{equation}
where the~$\nugr_\sigma\ldef\F^{\sigma\sigma}$ are called the coefficients of
expansion~\cite{Oseledec1968}.  The~$\nugr_\sigma$ give the instantaneous
relative growth of the principal axes of an infinitesimal ellipsoid moving
with the flow.  Taking the limit~$\time\rightarrow\infty$ in
Eq.~\eqref{eq:FTLEdef} gives the Lyapunov exponents~$\lyapexpinf_\sigma$,
which are independent of~$\lagrcv$ and~$\time$ in a given ergodic
region~\cite{Oseledec1968} (for almost all initial conditions); they converge
very slowly~\cite{Goldhirsch1987}.  \comment{Theorem 4, p. 213 of Oseledec.
What about the ``condition (5) of Sec. 2''?}  Associated with the coefficients
of expansion are the characteristic
directions~\hbox{$(\ediruv_\sigma)_q=\V_{q\sigma}$}, which converge
exponentially rapidly to their time-asymptotic
value~$\ediruvinf_\sigma(\lagrcv)$~\cite{Goldhirsch1987,Thiffeault2001b} (for
nondegenerate Lyapunov exponents).  They give the directions of stretching of
the axes of the ellipsoid in Lagrangian space.

Following Goldhirsch~\etal~\cite{Goldhirsch1987}, we shall assume that the
Lyapunov exponents~$\lyapexpinf_\sigma$ are nondegenerate and ordered such
that~\hbox{$\nugr_{\sigma-1} > \nugr_\sigma$}.  After allowing some time for
chaotic behaviour to set in, we have that~\hbox{$\nugr_\sigma \gg
\nugr_\kappa$} for~$\sigma<\kappa$.  Use will be made of this ordering in the
next section to estimate the asymptotic behaviour of Lagrangian derivatives.

\section{Lagrangian Derivatives}
\label{sec:LagrDerivs}

In Thiffeault~\cite{Thiffeault2001b}, evolution equations for the Lagrangian
derivatives of~$\ediruv$ ($\V$), $\wdirv$ ($\U$), and~$\nugr$ ($\F$) were
derived.  In this section we show how these equations must be modified to
account for a nontrivial metric.  We then demonstrate that the asymptotic form
for the Lagrangian derivatives, also obtained in Ref.~\cite{Thiffeault2001b},
applies to the curved case.

We define the quantities
\begin{equation}
	\Vgrnu_{\kappa\nu} \ldef {\Vu^p}{}_\kappa\,
		\frac{\pd}{\pd\lagrc^p}\log{\F^\nu}_\nu,
	\qquad
	\VgrU_{\kappa\mu\nu} \ldef {\Vu^p}{}_\kappa\,{\U^i}_\mu\,
		\covd_p{\U_{i\nu}},
	\label{eq:VgrUnudef}
\end{equation}
\begin{equation}
	\VgrV_{\kappa\mu\nu} \ldef
		{\Vu^p}{}_\kappa\,{\Vu^q}{}_\mu\,
		\frac{\pd\V_{q\nu}}{\pd\lagrc^p}
	\label{eq:VgrVdef}
\end{equation}
which are the Lagrangian derivatives of~$\nugr_\nu$,~$\U_{i\nu}$,
and~$\V_{q\nu}$ expressed in the orthonormal frames~$\{\wdirv_\sigma\}$
and~$\{\ediruv_\sigma\}$.  Integration by parts leads to the
symmetries~\hbox{$\VgrU_{\kappa\mu\nu} = -\VgrU_{\kappa\nu\mu}$}
and~\hbox{$\VgrV_{\kappa\mu\nu} = -\VgrV_{\kappa\nu\mu}$}.  We are using an
ordinary spatial derivative---as opposed to a covariant derivative---in the
definition of~$\VgrV$ because we have effectively rescaled the
metric~$\metrica_{pq}$ to give~$\eta_{pq}$ (Section~\ref{sec:curvedsp}).
Using the covariant derivative would destroy the antisymmetry of~$\VgrV$ in
its last two indices.

Equations of motion for~$\Vgrnu$,~$\VgrU$, and~$\VgrV$ can be derived in a
manner analogous to Ref.~\cite{Thiffeault2001b}, by taking Lagrangian
derivatives of the equations of
motion~\eqref{eq:SVDnugrODE}--\eqref{eq:SVDVODE}.  However, care must be taken
when commuting covariant derivatives: since the metric~$\metricx$ is
nontrivial, curvature terms can arise.  Moreover, a form of time-curvature
associated with the nontrivial time-dependence of~$\metricx$ is also
present~\cite{Thiffeault2001e}.  The evolution equation for~$\VgrU$ is
\begin{multline}
	\dot\VgrU_{\kappa\mu\nu} =
	\sum_\sigma{(\transp{\V}\dot\V)^\sigma}_\kappa\,\VgrU_{\sigma\mu\nu}
	+ \sum_\sigma{(\transp{\U}\Dt\U)^\sigma}_\mu\,\VgrU_{\kappa\sigma\nu}
	- \sum_\sigma{(\transp{\U}\Dt\U)^\sigma}_\nu\,\VgrU_{\kappa\mu\sigma}\\
		+ {\Vu}^q{}_\kappa\,
			\frac{\pd}{\pd\lagrc^q}(\transp{\U}\Dt\U)_{\mu\nu}
		- \nugr_\kappa\,(\sum_\sigma
			\Rcurv_{\mu\nu\kappa\sigma}\,\vel^\sigma
		+ \Scurv_{\mu\nu\kappa})
	\label{eq:VgrUODE}
\end{multline}
where the matrices~$(\transp{\U}\Dt\U)$ and~$(\transp{\V}\dot\V)$ denote the
right-hand side of Eqs. \eqref{eq:SVDUODE} and~\eqref{eq:SVDVODE},
respectively.  The last term in Eq.~\eqref{eq:VgrUODE} appears only
when~$\metricx$ has nontrivial spatial and temporal dependence.  It contains
the curvature~$\Rcurv$, defined by~\cite{Wald}
\begin{equation}
	\sum_\sigma\Rcurv_{\mu\nu\kappa\sigma}\,\vel^\sigma =
	{\U^k}_\mu\,{\U^i}_\nu\,{\U^j}_\kappa
	\l[\covd_k\,\covd_i\,\vel_j - \covd_i\,\covd_k\,\vel_j\r]\,,
	\label{eq:Rcurvdef}
\end{equation}
and the time-curvature~$\Scurv$~\cite{Thiffeault2001e}, defined by
\begin{equation}
	\Scurv_{\mu\nu\kappa} \ldef
	{\U^k}_\mu\,{\U^i}_\nu\,{\U^j}_\kappa
	\l[\covd_k\ptconnex_{ji} - \covd_i\ptconnex_{jk}\r]\,.
	\label{eq:Scurvdef}
\end{equation}
Both~$\Rcurv_{\mu\nu\kappa\sigma}$ and~$\Scurv_{\mu\nu\kappa}$ are
antisymmetric in~$\mu$ and~$\nu$.

Equations of motion can also be found for~$\Vgrnu$ and~$\VgrV$,
\begin{equation}
\dot\Vgrnu_{\kappa\nu} =
	\sum_\sigma{(\transp{\V}\dot\V)^\sigma}_\kappa\,\Vgrnu_{\sigma\nu}
	+ \sum_\sigma{\A^\sigma}_\nu\,\VgrU_{\kappa\sigma\nu}
	+ \nugr_\kappa\,\X_{\nu\kappa\nu}\,,
	\label{eq:VgrnuODE}
\end{equation}
\begin{multline}
	\dot\VgrV_{\kappa\mu\nu} =
	\sum_\sigma{(\transp{\V}\dot\V)^\sigma}_\kappa\,\VgrV_{\sigma\mu\nu}
	+ \sum_\sigma{(\transp{\V}\dot\V)^\sigma}_\mu\,\VgrV_{\kappa\sigma\nu}
	- \sum_\sigma{(\transp{\V}\dot\V)^\sigma}_\nu\,\VgrV_{\kappa\mu\sigma}
		\\
		+ {\Vu}^q{}_\kappa\,
			\frac{\pd}{\pd\lagrc^q}(\transp{\V}\dot\V)_{\mu\nu}\,,
	\label{eq:VgrVODE}
\end{multline}
where
\begin{equation}
	\X_{\nu\kappa\mu} \ldef
		{\U^k}_\nu\,{\U^i}_\kappa\,{\U^\ell}_\mu\,
		\covd_i\l[\covd_\ell\,\vel_k + \ptconnex_{k\ell}\r]\,.
	\label{eq:Xdef}
\end{equation}
Equations~\eqref{eq:VgrnuODE} and~\eqref{eq:VgrVODE} are identical in form to
their Euclidean versions of Ref.~\cite{Thiffeault2001e}, with the proviso that
the~$\widehat\X$ in that paper must be replaced by~$\X$ of
Eq.~\eqref{eq:Xdef}.  In its new form, which allows for a nontrivial metric,
$\X_{\nu\kappa\mu}$ is no longer symmetric in~$\kappa$ and~$\mu$; rather, it
satisfies the commutation relation
\begin{equation}
	\X_{\nu\kappa\mu} - \X_{\nu\mu\kappa} =
	\sum_\sigma\Rcurv_{\mu\nu\kappa\sigma}\,\vel^\sigma
	+ \Scurv_{\mu\nu\kappa}\,.
\end{equation}

Without redoing the entire derivation, we have thus shown that the evolution
equations derived in Ref.~\cite{Thiffeault2001b} can be easily modified to
allow for a nontrivial metric~$\metricx$.  Moreover, it is straightforward to
show that these terms do not modify the asymptotic behaviour of the Lagrangian
derivatives derived in Ref.~\cite{Thiffeault2001b}.  The reason is that if we
assume that the motion takes place on some bounded region of phase space, then
the curvatures~$\Rcurv$ and~$\Scurv$ must be bounded.

We now quote the result of Ref.~\cite{Thiffeault2001b}, which also applies to
our system: for \hbox{$\time\gg 1$}, by which we mean that the dynamical
system has evolved long enough for the quantities~$\nugr_\mu$ to have reached
a regime of quasi-exponential behaviour, we have that the Lagrangian
derivatives defined by Eqs.~\eqref{eq:VgrUnudef} and~\eqref{eq:VgrVdef} evolve
asymptotically as
\begin{align}
\VgrU_{\kappa\mu\nu} &= 
		\max\l(\nugr_\kappa,\nudec_{\mu\nu}\r)
		\VgrUt_{\kappa\mu\nu}\,,
	\label{eq:VgrUasym}\\
\Vgrnu_{\kappa\nu} &= 
		\max\l(\nugr_\kappa,\nudec_{\kappa\kappa},
		\nudec_{\nu\nu}\r)\Vgrnut_{\kappa\nu}
		+ \Vgrnuinf_{\kappa\nu}\,,
	\label{eq:Vgrnuasym}\\
\VgrV_{\kappa\mu\nu} &=
		\max\l(\nudec_{\mu\nu}\nugr_\kappa,
		\nudec_{\kappa\kappa},
		\nudec_{\mu\mu},
		\nudec_{\nu\nu}\r)
		\VgrVt_{\kappa\mu\nu}
		+ \VgrVinf_{\kappa\mu\nu}\,,
	\label{eq:VgrVasym}
\end{align}
where
\begin{equation}
	\nudec_{\mu\nu} \ldef
	\l\{\begin{array}{ll}
		{\nugr_\nu}/{\nugr_\mu}\,, & \mu < \nu;\\
		{\nugr_\mu}/{\nugr_\nu}\,, & \mu > \nu;\\
		\max\l(\frac{\nugr_{\mu}}{\nugr_{\mu-1}},
			\frac{\nugr_{\mu+1}}{\nugr_{\mu}}\r),\quad
			& \mu = \nu.
	\end{array}\r.
	\label{eq:nudecdef}
\end{equation}
The quantities with tildes have algebraic (nonexponential) behaviour, and the
quantities with~${}^\infty$ superscripts are constants.  The~$\nudec$'s are
defined such that asymptotically they are decreasing exponentially, for
nondegenerate~$\nugr$.  See Ref.~\cite{Thiffeault2001b} for a more detailed
discussion of the character of this asymptotic behaviour.

\section{Constraints}
\label{sec:constraints}

The Hessian~$\flowt_{**\lagrcv}$, introduced in~Section~\ref{sec:curvedsp}, is
the tensor of second derivatives of~$\flowt(\lagrcv)$; it characterises
deformations of fluid elements beyond ellipsoidal.  Because partial
derivatives commute, the Hessian is symmetric in the corresponding indices.
In Ref.~\cite{Thiffeault2001b} we showed that this symmetry of the Hessian
implies that the Lagrangian derivatives of Section~\ref{sec:LagrDerivs} are
not all independent.  However, the relationship between the derivatives is
very singular for chaotic flows.  We show in this section that the relations
can be used to obtain constraints on the derivatives, recovering and extending
previous results~\cite{Tang1996,Tang1999b,Thiffeault2001} deduced by different
means.

\subsection{Symmetry of the Hessian}
\label{sec:symmetry-hessian}

We define the projection of the Hessian~\hbox{$(\flowt_{**\lagrcv})^\ell_{pq}
= {\pd^2\x^\ell} / {\pd\lagrc^p\pd\lagrc^q}$} onto the~$\U$ and~$\V$ bases as
\begin{equation}
	\Hesm_{\kappa\mu\nu} \ldef
	\U_{\ell\kappa}\,(\flowt_{**\lagrcv})^\ell_{pq}\,
	{\Vu^p}{}_\mu\,{\Vu^q}{}_\nu\,  = \Hesm_{\kappa\nu\mu}\,.
	\label{eq:Hesmdef}
\end{equation}
Since~${(\flowt_{*\lagrcv})^\ell}_q=\pd\x^\ell/\pd\lagrc^q$, we can write
\begin{equation}
\frac{\pd^2\x^\ell}{\pd\lagrc^p\pd\lagrc^q}
		= \frac{\pd{(\flowt_{*\lagrcv})^\ell}_q}{\pd\lagrc^p},
\end{equation}
and using the decomposition~\eqref{eq:decomp}
for~${\M^i}_q={(\flowt_{*\lagrcv})^i}_q$, we find
\begin{equation}
	\Hesm_{\kappa\mu\nu}
	= \nugr_\kappa\,\Vgrnu_{\mu\kappa}\,\eta_{\nu\kappa}
		+ \nugr_\kappa\,\VgrV_{\mu\nu\kappa}
		+ \nugr_\nu\,\VgrU_{\mu\kappa\nu}\,.
	\label{eq:HesmSVD}
\end{equation}
The Hessian is symmetric in its lower indices, so we can equally well write
\begin{equation}
	\Hesm_{\kappa\mu\nu}
		= \nugr_\kappa\,\Vgrnu_{\nu\kappa}\,\eta_{\mu\kappa}
		+ \nugr_\kappa\,\VgrV_{\nu\mu\kappa}
		+ \nugr_\mu\,\VgrU_{\nu\kappa\mu}\,,
	\label{eq:HesmSVDsym}
\end{equation}
where we interchanged~$\mu$ and~$\nu$ in~\eqref{eq:HesmSVD}.
Equating~\eqref{eq:HesmSVD} and~\eqref{eq:HesmSVDsym}, we find the relations
\begin{alignat}{2}
	\nugr_\mu\l(\VgrV_{\mu\mu\nu} + \eta_{\mu\mu}\Vgrnu_{\nu\mu}\r)
		&= \nugr_\nu\,\VgrU_{\mu\mu\nu}\,,
		&&\mu\ne\nu,
	\label{eq:Hessymrel1}\\
	\nugr_\kappa\l(\VgrV_{\mu\nu\kappa} - \VgrV_{\nu\mu\kappa}\r)
		&= \nugr_\nu\,\VgrU_{\mu\nu\kappa} -
		\nugr_\mu\,\VgrU_{\nu\mu\kappa}\,,
		&\qquad& \mu,\nu,\kappa\ \mathrm{differ}.
	\label{eq:Hessymrel2}
\end{alignat}
Equations~\eqref{eq:Hessymrel1} and~\eqref{eq:Hessymrel2} can be used to solve
for the~$\VgrV$'s in terms of the~$\VgrU$'s and~$\Vgrnu$'s.  However, for
chaotic flows the exponential behaviour of the~$\nugr$'s renders this solution
highly singular~\cite{Thiffeault2001b}.  Thus, in practise it is better to
evolve all the Lagrangian derivatives.

Rather than solving for the~$\VgrV$, if the flow is chaotic the
relations~\eqref{eq:Hessymrel1} and~\eqref{eq:Hessymrel2} can be put to good
use in another manner.  The Lagrangian derivatives of~$\U$, as contained
in~$\VgrU$, are not quantities of great interest to us.  They contain
information about the sensitive dependence on initial conditions of the
absolute orientation of fluid elements in Eulerian space.  This information is
not necessary for solving problems in Lagrangian coordinates, and is too
sensitive to initial conditions to be of use anyhow.  We thus substitute the
time-asymptotic form of~$\VgrU$, given by Eq.~\eqref{eq:VgrUasym}, in the
right-hand side of Eqs.~\eqref{eq:Hessymrel1} and~\eqref{eq:Hessymrel2},
yielding
\begin{align}
	\VgrV_{\mu\mu\nu} + \eta_{\mu\mu}\Vgrnu_{\nu\mu}
		&= \max\l(\nugr_\nu\,,\frac{\nugr_\nu}{\nugr_\mu}
		\,\,\nudec_{\mu\nu}\r)\VgrUt_{\mu\mu\nu}\,,
	\label{eq:aconstrtype1}\\
	\VgrV_{\mu\nu\kappa} - \VgrV_{\nu\mu\kappa}
		&= \frac{\nugr_\nu}{\nugr_\kappa}\,
			\max\l(\nugr_\mu\,,\nudec_{\nu\kappa}\r)
			\VgrUt_{\mu\nu\kappa}
		- \frac{\nugr_\mu}{\nugr_\kappa}\,
			\max\l(\nugr_\nu\,,\nudec_{\mu\kappa}\r)
			\VgrUt_{\nu\mu\kappa}\,,
	\label{eq:aconstrtype2}
\end{align}
where, as before,~$\mu$, $\nu$, and~$\kappa$ differ.

\subsection{Type I Constraints}
\label{sec:type1constr}

For~$\mu<\nu$, Eq.~\eqref{eq:aconstrtype1} can be written
\begin{equation}
	\VgrV_{\mu\mu\nu} + \Vgrnu_{\nu\mu}
		= \max\l(\nugr_\nu\,,\,\nudec_{\mu\nu}^2\r)
		\VgrUt_{\mu\mu\nu}\,,
	\qquad \mu<\nu,
	\label{eq:constrtype1}
\end{equation}
where we have set~$\eta_{\mu\mu}=1$ for a proper Riemannian metric.  If the
index~$\nu$ corresponds to a contracting direction ($\nugr_\nu\ll1$), the
right-hand side of Eq.~\eqref{eq:constrtype1} goes to zero exponentially fast,
at a rate~$\nugr_\nu$ or~$\nudec_{\mu\nu}^2$, whichever is slowest.  In that
case Eq.~\eqref{eq:constrtype1} is a \emph{constraint} implying that for
large~$\time$ we have
\begin{equation}
	(\VgrV_{\mu\mu\nu} + \Vgrnu_{\nu\mu}) \longrightarrow 0,
	\qquad \nugr_\nu\ll\nugr_\mu\,,\ \ \nugr_\nu\ll 1.
	\label{eq:constrtype1b}
\end{equation}
We refer to~\eqref{eq:constrtype1b} as Type I constraints.  The total number of
such Type I constraints is
\begin{equation}
	\NI = \ns\l[\sdim-\half(\ns+1)\r],
	\label{eq:NI}
\end{equation}
where~$\sdim$ is the dimension of the space and~$\ns$ is the number of
contracting directions (i.e., the number of negative Lyapunov exponents)
possessed by the flow in a particular chaotic region.  Table~\ref{tab:NI}
gives the number of Type I constraints,~$\NI$, as a function of~$\sdim$
and~$\ns$.
\begin{table}
\caption{The total number of Type I constraints for low-dimensional systems,
as given by Eq.~\eqref{eq:NI}.  The rows denote~$\sdim$, the
columns~$\ns\le\sdim$.}
\label{tab:NI}
\vspace{1em}
\begin{center}
\begin{tabular}{ccccccccc}
\hline
$\sdim$  && \textsl{1} & \textsl{2} & \textsl{3} & \textsl{4} & \textsl{5} &
\textsl{6} & \textsl{7}\\
\hline
\textsl{1} &\quad& 0 & & & & & &\\
\textsl{2} && 1 & 1 &  & & & &\\
\textsl{3} && 2 & 3 & 3 & & & &\\
\textsl{4} && 3 & 5 & 6 & 6 & & &\\
\textsl{5} && 4 & 7 & 9 & 10 & 10 & &\\
\textsl{6} && 5 & 9 & 12 & 14 & 15 & 15 &\\
\textsl{7} && 6 & 11 & 15 & 18 & 20 & 21 & 21\\
\hline
\end{tabular}
\end{center}
\vspace{1em}
\end{table}

In two dimensions, we typically have one contracting direction, so there is a
single Type I constraint.  This is the same constraint that was derived in
Refs.~\cite{Tang1996,Thiffeault2001}.

In three dimensions, for an autonomous flow, we typically also have one
contracting direction.  There are then two Type I constraint.  These
constraints correspond to those derived in Ref.~\cite{Thiffeault2001}.

An interesting special case of the Type I constraints is obtained by
setting~\hbox{$\nu=\sdim$} in Eq.~\eqref{eq:constrtype1}, and then summing
over~$\mu<\sdim$.  We obtain
\begin{equation}
	\covd_q\,(\ediruv_n)^q - (\ediruv_n)^q\,\covd_q\log\nugr_\sdim
	\sim \max\l(\nugr_\sdim\,,\,\nudec_{\sdim\sdim}^2\r)
	\rightarrow 0.
	\label{eq:divsconstr}
\end{equation}
\comment{Verify this.  Mention covariant derivative: okay to use here because
the exponential factor in the determinant of $\metrica$ cancels out.}  This
constraint was discovered numerically and used by Tang and
Boozer~\cite{Tang1996,Tang1999b,Tang1999a,Tang2000} and derived in three
dimensions by Thiffeault and Boozer~\cite{Thiffeault2001}.  The present method
not only gives the constraint in a very direct manner for any
dimension~$\sdim$, but it also provides us with its asymptotic convergence
rate, as determined by the right-hand side of~\eqref{eq:divsconstr}.

\subsection{Type II Constraints}
\label{sec:type2constr}

Equation~\eqref{eq:constrtype2} implies that
\begin{equation}
	\VgrV_{\mu\nu\kappa} - \VgrV_{\nu\mu\kappa}
		\sim
		\max\l(\frac{\nugr_\mu\,\nugr_\nu}{\nugr_\kappa}\,,
		\frac{\nugr_\nu}{\nugr_\kappa}\,\nudec_{\nu\kappa}\,,
		\frac{\nugr_\mu}{\nugr_\kappa}\,\nudec_{\mu\kappa}\r).
	\label{eq:aaconstrtype2}
\end{equation}
We are interested in finding constraints analogous to the Type I constraints
of Section~\ref{sec:type1constr}.  It is clear that unless both~$\mu$
and~$\nu$ are greater than~$\kappa$, the right-hand side of
Eq.~\eqref{eq:aaconstrtype2} is of order unity or greater, and so does not go
to zero.  We can assume without loss of generality that~$\mu<\nu$, so that
\begin{equation}
	\VgrV_{\mu\nu\kappa} - \VgrV_{\nu\mu\kappa} \sim
		\nudec_{\mu\kappa}\,\max\l(\nugr_\nu\,,
		\nudec_{\mu\kappa}\r),\qquad
	\nu>\mu>\kappa,
	\label{eq:constrtype2}
\end{equation}
where we have used~$\nudec_{\nu\kappa}\ll\nudec_{\mu\kappa}$.  Whether or not
Eq.~\eqref{eq:constrtype2} is a constraint depends on the specific behaviour
of~$\nugr_\mu\,\nugr_\nu/\nugr_\kappa$.  Clearly, we have a constraint
if~\hbox{$\nugr_\nu\ll 1$}, since~\hbox{$\nudec_{\mu\kappa}\ll 1$}.  This
provides a lower bound on the number~$\NII$ of Type II constraints; by
choosing~$\nu$ from the~$\ns$ contracting directions, and summing over the
remaining~\hbox{$\kappa>\mu>\nu$}, we obtain
\begin{equation}
	\NII \ge
	\half\ns\l[\sdim^2 - (\ns + 2)\l(\sdim - \third(\ns + 1)\r)\r].
	\label{eq:NIImin}
\end{equation}
But even if~\hbox{$\nugr_\nu\gg1$} we can have a constraint, as long
as~\hbox{$\nugr_\nu\,\nudec_{\mu\kappa}\ll1$}.  This depends on the particular
problem at hand; hence, Eq.~\eqref{eq:NIImin} is only a lower bound, but a
fairly tight one for low dimensions.  Table~\ref{tab:NIImin} enumerates the
minimum number of Type II constraints as a function of~$\sdim$ and~$\ns$.
\begin{table}
\caption{Lower bound on the number of Type II constraints, as given by
Eq.~\eqref{eq:NIImin}.  The rows denote~$\sdim$, the columns~$\ns\le\sdim$.}
\label{tab:NIImin}
\vspace{1em}
\begin{center}
\begin{tabular}{ccccccccc}
\hline
$\sdim$ && \textsl{1} & \textsl{2} & \textsl{3} & \textsl{4} & \textsl{5} &
\textsl{6} & \textsl{7}\\
\hline
\textsl{1} &\quad& 0 & & & & & &\\
\textsl{2} && 0 & 0 &  & & & &\\
\textsl{3} && 1 & 1 & 1 & & & &\\
\textsl{4} && 3 & 4 & 4 & 4 & & &\\
\textsl{5} && 6 & 9 & 10 & 10 & 10 & &\\
\textsl{6} && 10 & 16 & 19 & 20 & 20 & 20 &\\
\textsl{7} && 15 & 25 & 31 & 34 & 35 & 35 & 35\\
\hline
\end{tabular}
\end{center}
\vspace{1em}
\end{table}

Note that when~$\kappa$, $\mu$, and $\nu$ differ we can write
\begin{equation}
	\VgrV_{\mu\nu\kappa} - \VgrV_{\nu\mu\kappa} =
		-(\ediruv_\kappa)_q\,\l[\ediruv_\mu\,,\ediruv_\nu\r]^q,
\end{equation}
where the Lie bracket is
\begin{equation}
	\l[\ediruv_\mu\,,\ediruv_\nu\r]^q \ldef
	(\ediruv_\mu)^p\,\frac{\pd}{\pd\lagrc^p}\,(\ediruv_\nu)^q
	- (\ediruv_\nu)^p\,\frac{\pd}{\pd\lagrc^p}\,(\ediruv_\mu)^q\,.
	\label{eq:Liebrakdef}
\end{equation}
The Type II constraints are thus forcing certain Lie brackets of the
characteristic directions~$\ediruv_\sigma$ to vanish asymptotically.  The
geometrical implications of this, and perhaps a connexion to the Frobenius
theorem and the existence of submanifolds, remains to be explored.

\section{Curvature}
\label{sec:curvature}

We now make contact with the earlier results of
Refs.~\cite{Tang1996,Thiffeault2001}, where the constraints were derived for
two and three dimensional flows by examining the form of the Riemann curvature
tensor associated with the metric~$\metrica$.  We list the advantages of the
present method.  The only known disadvantage is that the formalism presented
here does not apply to maps, only to continuous flows.  This is because of the
lack of an \SVD\ method for maps~\cite{Geist1990}.  The use of the \QR\
method, which is defined for maps\cite{Eckmann1985,Geist1990}, may circumvent
that problem, but the analysis is more difficult and has not been carried out
yet.

Nontrivial metrics can have curvature, as reflected by the presence of the
Riemann curvature tensor in Section~\ref{sec:LagrDerivs}.  A straightforward
method of obtaining that tensor is through the use of \emph{Ricci rotation
coefficients}~\cite{Wald},
\begin{equation}
	\Riccirotc_{\kappa\mu\nu} \ldef
	(\wdir_\kappa)^i\,(\wdir_\mu)^j\,\covd_i\,(\wdir_\nu)_j
	= (\edir_\kappa)^p\,(\edir_\mu)^q\,\covd_p\,(\edir_\nu)_q\,.
\end{equation}
These satisfy the antisymmetry property~\hbox{$\Riccirotc_{\kappa\mu\nu} =
-\Riccirotc_{\kappa\nu\mu}$}, and can be rewritten in terms of the~$\VgrU$ of
Eq.~\eqref{eq:VgrUnudef} as
\begin{equation}
	\Riccirotc_{\kappa\mu\nu} = \nugr_\kappa^{-1}\,\VgrU_{\kappa\mu\nu}\,.
	\label{eq:rotcVgrU}
\end{equation}
In terms of the rotation coefficients, the Riemann curvature tensor
is~\cite[p.~51]{Wald}
\begin{multline}
	\Rcurv_{\mu\nu\kappa\sigma} =
	(\edir_\kappa)^q\,\covd_q\,\Riccirotc_{\sigma\mu\nu}
	- (\edir_\sigma)^q\,\covd_q\,\Riccirotc_{\kappa\mu\nu}\\
	- \eta^{\tau\rho}\l[
	\Riccirotc_{\kappa\rho\mu}\,\Riccirotc_{\sigma\tau\nu}
	- \Riccirotc_{\sigma\rho\mu}\,\Riccirotc_{\kappa\tau\nu}
	+ \Riccirotc_{\kappa\rho\sigma}\,\Riccirotc_{\tau\mu\nu}
	- \Riccirotc_{\sigma\rho\kappa}\,\Riccirotc_{\tau\mu\nu}
	\r].
	\label{eq:Rcurvrotc}
\end{multline}
If we use the relations~\eqref{eq:Hessymrel1} and~\eqref{eq:Hessymrel2} to
solve for~$\VgrU$ in terms of~$\VgrV$ and~$\Vgrnu$, we can rewrite
Eq.~\eqref{eq:rotcVgrU} as
\begin{multline}
	\Riccirotc_{\kappa\mu\nu} =
	\frac{1}{2\nugr_\kappa\nugr_\mu\nugr_\nu}
	\l\{\nugr_\mu^2(\VgrV_{\nu\kappa\mu} - \VgrV_{\kappa\nu\mu})
	+ \nugr_\nu^2(\VgrV_{\kappa\mu\nu} - \VgrV_{\mu\kappa\nu})
	- \nugr_\kappa^2(\VgrV_{\mu\nu\kappa} - \VgrV_{\nu\mu\kappa})\r\}\\
	+ \frac{1}{\nugr_\nu}\,\eta_{\mu\kappa}\,\Vgrnu_{\nu\mu}
	- \frac{1}{\nugr_\mu}\,\eta_{\nu\kappa}\,\Vgrnu_{\mu\nu}
	\label{eq:rotcVgrVnu}
\end{multline}
The form of the curvature obtained by inserting Eq.~\eqref{eq:rotcVgrVnu} into
\eqref{eq:Rcurvrotc} is essentially the one obtained in three dimensions in
Ref.~\cite{Thiffeault2001}.  The curvature was calculated directly from the
Christoffel symbols in Ref.~\cite{Tang1996}.  The constraints were then
deduced by imposing the finiteness of the curvature tensor: some terms in
Eq.~\eqref{eq:rotcVgrVnu} would appear to grow exponentially, so their
coefficient must go to zero to maintain the curvature finite.  In this manner,
the Type I and Type II constraints were derived in two and three dimensions,
backed by numerical evidence~\cite{Tang1996,Thiffeault2001,Thiffeault2001b}.

The approach used in this paper to derive the constraints is advantageous in
several ways: (i) It is valid in any number of dimensions; (ii) It avoids
using the curvature; (iii) There are no assumptions about the growth rate of
individual terms in the curvature~\cite{Thiffeault2001}; (iv) The convergence
rate of the constraints is rigorously obtained [Eqs.~\eqref{eq:constrtype1}
and~\eqref{eq:constrtype2}]; (v) The number of constraints can be predicted
[Eqs.~\eqref{eq:NI} and~\eqref{eq:NIImin}].  The crux of the difference
between the two approaches is that here we use the variable~$\VgrU$ to
estimate the asymptotic behaviour of the constraints directly, rather than
relying on indirect evidence from the curvature.  Thus, the derivation of the
time-asymptotic form of~$\VgrU$ is essential (Section~\ref{sec:LagrDerivs} and
Ref.~\cite{Thiffeault2001b}).

\section{Discussion}
\label{sec:discussion}

In this paper we have demonstrated that in chaotic flows the Lagrangian
derivatives of finite-time Lyapunov exponents, and their associated
characteristic directions, satisfy time-asymptotic differential constraints.
The constraints are valid for any metric tensor, and are satisfied with
quasi-exponential (that is, exponential in the long-time limit) accuracy in
time.  The method employed to derive the constraints made use of earlier work
on the asymptotic form of Lagrangian derivatives~\cite{Thiffeault2001b}.  The
new results generalise previous work~\cite{Tang1996,Thiffeault2001} by
providing constraints in any number of dimensions, together with their
convergence rate.  We now give a few examples of possible applications of the
constraints, and directions for future research.

In Ref.~\cite{Tang1996} it was demonstrated that for the isotropic, homogeneous
advection--diffusion equation the diffusion proceeds almost exclusively along
the characteristic direction associated with the smallest Lyapunov exponents.
This was shown by transforming the equation to Lagrangian coordinates, thereby
eliminating the advection term.  The resulting equation was an anisotropic
diffusion equation, where the diffusion tensor was proportional
to~$\metrica^{-1}$, the inverse of the metric tensor in Lagrangian space.
Because the inverse of the metric appears, the dominant eigendirection is that
associated with the \emph{smallest} Lyapunov exponent.  Physically, this means
that diffusion occurs in the direction along which large gradients are created
by the stretching effect of the flow.

If the diffusion tensor in the advection--diffusion equation is anisotropic
and inhomogeneous, then the direction of fastest diffusion is associated with
the characteristic direction calculated using the diffusion tensor itself as a
metric.  Furthermore, in the large P\'{e}clet number limit (small
diffusivity), the Type I constraints derived in Section~\ref{sec:type1constr}
can actually be used to transform the advection--diffusion equation into a
scalar one-dimensional diffusion equation.  This use of the constraints will
be addressed in subsequent work~\cite{Thiffeault2001f}.  Other applications of
the Type I constraint can be found in Tang and
Boozer~\cite{Tang1999a,Tang2000}: they demonstrate that regions of high
curvature of the integral curves of~$\ediruv_\sdim$ lead to locally small
finite-time Lyapunov exponents, hindering the chaotic enhancement of
diffusion.

In three dimensions with an Euclidean metric, the Type II
constraint~\eqref{eq:constrtype2} can be written
\begin{equation}
	\ediruv_1\cdot\curl\ediruv_1 \sim
	\nudec_{12}\,\max\l(\nugr_3\,,
		\nudec_{12}\r) \rightarrow 0,
	\label{eq:udotcurlu}
\end{equation}
where the curl is taken with respect to Lagrangian coordinates,~$\lagrcv$.
Equation~\eqref{eq:udotcurlu} implies that, asymptotically,~$\ediruv_1$ is of
the form
\begin{equation}
	\ediruv_1 = \grad\ppot/\norm{\grad\ppot},
	\label{eq:uppot}
\end{equation}
where~$\norm{\cdot}$ is the Euclidean norm, and~$\ppot$ is a pseudopotential
function.  Equation~\eqref{eq:uppot} is valid locally.  Hence, the field of
unstable directions (associated with the largest Lyapunov exponent) has a
simple asymptotic structure.  The implications of that structure remain to be
investigated.  An indication of how it might prove useful was investigated in
Ref.~\cite{Thiffeault2001}, where it was shown that the Type II constraints
could have an impact on energy dissipation in the fast kinematic dynamo.

\begin{ack}

The author thanks Allen H. Boozer for helpful discussions.  This work was
supported by the National Science Foundation and the Department of Energy
under a Partnership in Basic Plasma Science grant, No.~DE-FG02-97ER54441.

\end{ack}


\appendix

\section{Singular Value Decomposition with a Metric}
\label{sec:SVDmetric}

The decomposition given by Eq.~\eqref{eq:decomp} is defined for a space with a
nontrivial metric, but the numerical algorithms for obtaining the \SVD\ are
based on ``true'' orthogonal matrices (that is, matrices orthogonal with
respect to the Euclidean metric).  In this appendix we show that we can always
find the bases~$\{\edirv_\sigma\}$ and~$\{\wdirv_\sigma\}$ using the ordinary
singular value decomposition of matrices with a Euclidean metric.

Since the matrix~$\metricx_{ij}$ of components of the
metric~$(\cdot,\cdot)_\xv$ on~$\Tangent_{\xv}\Manif$ is symmetric, it can be
diagonalised by an orthogonal transformation.  This transformation can then be
composed with a rescaling to make the diagonal elements equal to unity (a
proper Riemannian metric has positive eigenvalues), yielding
\begin{equation}
	\metricx_{ij} = {\dsq^{i'}}_i\,\eta_{i'j'}\,
		{\dsq^{j'}}_j,
	\label{eq:dsqdef}
\end{equation}
where~${\dsq^{i'}}_i(\time,\xv)$ is the coordinate transformation,
and~$\eta_{i'j'}=\delta_{i'j'}$ for a positive-definite metric.  We are using
primed indices to denote this new basis on~$\Tangent_{\xv}\Manif$
where~$\metricx$ is diagonal.  The metric~$\metrica$ of
Eq.~\eqref{eq:metricacomp} can then be written
\begin{equation}
	\metrica_{pq}
	= {(\flowt_{*\lagrcv})^i}_p\,
		{\dsq^{i'}}_i\,\eta_{i'j'}\,{\dsq^{j'}}_j\,
		{(\flowt_{*\lagrcv})^j}_q\,.
	\label{eq:metricadsqeta}
\end{equation}
Define the matrix~$\dM$ with components
\begin{equation}
	{\dM^{i'}}_q \ldef {\dsq^{i'}}_k\,{(\flowt_{*\lagrcv})^k}_q\,,
	\label{eq:dMdef}
\end{equation}
so that Eq.~\eqref{eq:metricadsqeta} can be written
\begin{equation}
	\metrica_{pq}
	= {\dM^{i'}}_p\,\eta_{i'j'}\,{\dM^{j'}}_q\,.
	\label{eq:metricadM}
\end{equation}
Since~$\eta_{i'j'}=\delta_{i'j'}$, the metric is now in a
form~$\metrica=\transp{\dM}\dM$, suggesting the use of the usual \SVD\
technique to eliminate the Eulerian information in~$\metrica$.  We have
absorbed the non-Euclidean metric into~$\dM$, so that even though the space is
not Euclidean the decomposition carries through in the same manner as in
Refs.~\cite{Greene1987,Thiffeault2001b}.

Proceeding with the singular value decomposition, we write
\begin{equation}
	{\dM^{i'}}_q = {\Q^{i'}}_\sigma\,
		\F^{\sigma\tau}\,\V_{q\tau}\,,
	\label{eq:SVD}
\end{equation}
where~$\Q$ and~$\V$ are ordinary orthogonal matrices (\ie, with respect to the
Euclidean metric) and~$\F$ is diagonal.

The basis~$\{\edirv_\sigma\}$ is given by~$\V$ and~$\F$ as in
Eq.~\eqref{eq:edirdef}.  The basis~$\{\wdir_\sigma\}$ is given in terms
of~$\Q$ and~$\dsq$ by
\begin{equation}
	(\wdirv_\sigma)^i = {\U^i}_\sigma =
		{{(\dsq^{-1})}^i}_{i'}\,{\Q^{i'}}_\sigma\,,
	\qquad
	(\wdirv_\sigma)_i = \U_{i\sigma} = \eta_{i'j'}\,
		{\dsq^{i'}}_{i}\,{\Q^{j'}}_\sigma\,.
\end{equation}
Since the \SVD\ always exists, we have demonstrated that the construction of
the basis vectors~$\edirv_\sigma$ and~$\wdirv_\sigma$ satisfying
Eqs.~\eqref{eq:flowtdecomp} and~\eqref{eq:metricdiag} is possible.

\end{document}